\documentclass[12pt]{spieman}  % 12pt font required by SPIE;
\usepackage{amsmath,amsfonts,amssymb}
\usepackage{graphicx}
\usepackage{setspace}
\usepackage{fancyhdr}

\title{The STROBE-X Wide Field Monitor Instrument}

\author[a,*]{Ronald A. Remillard}
\author[b]{Margarita Hernanz}
\author[c]{Jean in 't Zand}
\author[d]{Paul S. Ray}
\author[e]{Valter Bonvicini}
\author[f]{Søren Brandt}
\author[c]{Terri Brandt}
\author[b]{Alex Carmona}
\author[g]{Yuri Evangelista}
\author[b]{Daniel Alvarez Franco}
\author[h]{Cynthia Froning}
\author[b]{José-Luis Galvez}
\author[d,o,p]{Gianluigi De Geronimo}
\author[c]{Martin Grim}
\author[i]{Emrah Kalemci}
\author[c]{Lucien Kuiper}
\author[f]{Irfan Kuvvetli}
\author[j]{Thomas J. Maccarone}
\author[k]{Witold Nowosielski}
\author[a]{Dheeraj R. R. Pasham}
\author[b]{Alessandro Patruno}
\author[h]{Steven C. Persyn}
\author[h]{Peter W. A. Roming}
\author[l]{Andrea Santangelo}
\author[m]{Stephane Schanne}
\author[l]{Christoph Tenzer}
\author[n]{Colleen A. Wilson-Hodge}
\author[e]{Gianluigi Zampa}
\author[c]{Frans Zwart}

\affil[a]{Kavli Institute for Astrophysics \& Space Research, Massachusetts Institute of Technology, 77 Massachusetts Avenue, Cambridge, MA 02139 USA }
\affil[b]{Institute of Space Sciences (ICE, CSIC) and IEEC, Campus UAB, Camí de Can Magrans s/n, 08193 Cerdanyola del Vallès (Barcelona), Spain}
\affil[c]{SRON Netherlands Institute fo<<r Space Research, Niels Bohrweg 4, 2333 CA Leiden, The Netherlands}
\affil[d]{Space Science Division, U.S. Naval Research Laboratory, Washington, DC 20375, USA}
\affil[e]{INFN Trieste, 34127 Trieste, Italy}
\affil[f]{DTU-Space, Technical University of Denmark, Lyngby, Denmark}
\affil[g]{University of Michigan, Nuclear Engineering and Radiological Sciences, Ann Arbor, MI 48109, USA}
\affil[h]{Southwest Research Institute, San Antonio, TX 78238 USA}
\affil[i]{Sabanci University, Faculty of Engineering and Natural Sciences, 34956, Istanbul, Turkey}
\affil[j]{Department of Physics \& Astronomy, Texas Tech University, Lubbock, TX, 79409, USA}
\affil[k]{CBK, Space Research Center, Polish Academy of Sciences, PL-00-716 Warszawa, Poland}
\affil[l]{IAAT Universit\"{a}t T\"{u}bingen, 72076 T\"{u}bingen, Germany}
\affil[m]{IRFU, CEA, Université Paris-Saclay, 91191 Gif-sur-Yvette, France}
\affil[n]{NASA/MSFC, Huntsville, AL, USA}
\affil[o]{Stony Brook University, Electrical and Computer Engineering, Stony Brook, NY 11794, USA}
\affil[p]{DG Circuits, dgcircuits.com, Syosset, NY 11973, USA}

\begin{document} 
\maketitle

\begin{abstract}
The Wide Field Monitor (WFM) is one of the three instruments on the
\textit{Spectroscopic Time-Resolving Observatory for Broadband Energy
  X-rays} (STROBE-X) mission, which was proposed in response to NASA's
2023 call for a probe-class mission.  The WFM is a coded-mask camera
system that would be the most scientifically capable wide-angle
monitor ever flown. The WFM will anchor X-ray time domain astronomy,
at the all-sky level, for the 2030s.  The field of view covers
one-third of the sky, to 50\% mask coding, and the energy sensitivity
is 2--50 keV. The WFM will identify new X-ray transients for rapid
observations with the two pointed instruments of STROBE-X.  In
addition, the WFM will capture spectral/timing changes in known
sources with data of unprecedented quality. WFM data will uniquely
advance scientific knowledge for diverse classes in high-energy
astrophysics, including X-ray bursts that coincide with gravitational
wave detections, gamma ray bursts and their transition from prompt
emission to afterglow, subluminous GRBs that may signal shock breakout
in supernovae, state transitions in accreting compact objects and
their jets, bright flares in fast X-ray transients, accretion onset in
transitional pulsars, and coronal flares from many types of active
stars.
\end{abstract}

% Include a list of up to six keywords after the abstract

% Include email contact information for corresponding author
{\noindent \footnotesize\textbf{*}\linkable{rr@space.mit.edu} }

\begin{spacing}{1}   % use double spacing for rest of manuscript

\section{Instrument Overview}
\label{sect:intro}  % \label{} allows reference to this section
We present an overview of the WFM Instrument on the STROBE-X Probe\cite{JATISOverview}, which was proposed to NASA in 2023.
The WFM cameras \cite{2022SPIE12181E..1YH} image the X-ray sky by placing position-sensing X-ray detectors below a coded mask, which is a metal plate perforated with rectangular slits \cite{Goldwurm2022}. Each X-ray source in the field of view (FoV) casts a shadow of the mask onto the detector plane. Deconvolution of the overlapping shadow patterns reveals the intensity of each contributing source.  New X-ray transients are recognized as post-deconvolution residuals that can be back-projected, via the mask, to a specific location on the sky. Coded mask cameras have been flown by NASA, ESA, and other space agencies, since the 1970s.  

Four pairs of coded-aperture cameras are deployed, with each pair providing 2-D images with a half-coded field of view (FoV) of $65^\circ \times 65^\circ$. The combined instantaneous FoV, to the half-coded limit, is 4 sr or 32\% of the sky. Adjacent camera pairs are mounted with a separation angle of $65^\circ$, which aligns their half-coded response and flattens the effective area across the central WFM FoV. The total FoV to the limit of zero response through the mask covers 56\% of the sky.

The WFM detectors employ the same silicon drift detector (SDD) technology \cite{2022hxga.book...68V} as one of the pointed instruments of STROBE-X, the High Energy Modular Array \cite{JATISHEMA} (HEMA), but with finer anode pitch to enhance imaging. The nominal energy band is 2--50 keV, with energy resolution better than 300 eV at 6 keV (see Fig.~4 in Ref.~\citenum{2012SPIE.8443E..5PE}). The source localization accuracy is 1 arcmin or better, depending on the source signal-to-noise ratio.  A distinguishing feature is that all events are telemetered to the ground with 24 $\mu$s time resolution, giving maximum flexibility for analyses with high spectral and timing resolution.

Multi-band imaging analyses and transient searches are performed onboard, allowing the WFM to issue prompt alerts to the community and to trigger autonomous slew requests to the spacecraft to observe the target of opportunity (ToO) with the STROBE-X pointed instruments. The goals are to issue community alerts and slew requests within 30 s of data acquisition, while source acquisition with {HEMA and the Low Energy Modular Array \cite{JATISLEMA} (LEMA) could happen within minutes, depending on the duration of the slew. 

The WFM cameras for STROBE-X are an adaptation of a design led by groups affiliated with the European Space Agency (ESA). The WFM team for STROBE-X includes many leaders of this effort, in addition to investigators who have worked on wide-angle X-ray instruments supported by NASA.  The Sections below describe the WFM design heritage, science overview, instrument components, operations on STROBE-X, calibration, and instrument performance. Simulations of the WFM have been a foundation for estimating end-to-end performance, including instrument response, image reconstruction, and sensitivity limits \cite{2012SPIE.8443E..5PE,2024Ceraudo...preprint}. The modular design of this software facilitates adaptations for the different configurations of described in the next Section. It is expected that the tools will be maintained, as the WFM for STROBE-X completes its definition, contributing directly to the software needed for the WFM analysis pipelines and the tools made available to the scientific community.

\section{Instrument Design Heritage}
The primary elements of the WFM design were first developed for the Large Observatory for X-ray Timing (LOFT) \cite{2014SPIE.9144E..2VB}, and a detailed description of WFM-LOFT was included in the ESA Assessment Study Report (“Yellow Book”) for LOFT \cite{LOFTYellowBook}.  Modest design revisions were made for the possibility of a WFM instrument on the enhanced X-ray Timing and Polarimetry mission (eXTP), under consideration for launch by China. WFM-eXTP is described and illustrated in Refs.~\citenum{2022SPIE12181E..1YH, 2022SPIE12181E..65G, 2022SPIE12181E..6FX, 2022SPIE12181E..67Z, 2022SPIE12181E..66K}. 

The collaboration between European and US teams for WFM-STROBE-X reached a formal basis with definition studies of STROBE-X by the Instrument Design Lab at Goddard Space Flight Center (GSFC) in 2017 and the Mission Design Lab at GSFC in 2018. This effort progressed into the NASA Probe proposal, submitted in 2023. The WFM-STROBE-X co-investigators and collaborators include members of instrument teams that conducted successful missions for wide-angle X-ray astronomy with, e.g., the Rossi X-ray Timing Explorer All-Sky Monitor (NASA) \cite{1996ApJ...469L..33L}, the BeppoSAX Wide Field Cameras (Netherlands/Italy) \cite{1997A&AS..125..557J}), the Swift Burst Alert Telescope (NASA) \cite{2004ApJ...611.1005G}, SuperAGILE (Italy) \cite{2007NIMPA.581..728F}, and the Fermi Burst Monitor (NASA) \cite{2009ApJ...702..791M}.

\section{Science Overview}
\label{sect:sciintro}  % \label{} allows reference to this section
Scientific contributions by the WFM can be described in terms of two broad themes: new transient discoveries and special opportunities for known X-ray sources.  The WFM will discover transient X-ray sources that are distributed over many source classes and a wide range of activity timescales.  Brief events, from ms to hours, will be captured and positioned on the sky for sources such as Gamma Ray Bursts (GRBs), related bursts associated with gravitational waves, so-called "fast X-ray transients", magnetar flashes, and isolated type I X-ray bursts. The WFM will also discover the onset of high-energy activity in new transients that will remain active from days to years.  The principal example, here, is the appearance of an X-ray binary system, where there is an episode of accretion onto a black hole (BH) or neutron star (NS). New sources are extremely important in high-energy astrophysics.  GRBs and related GW counterparts are transient and non-recurring, by their nature. Studies of black hole binaries, rely heavily on transient sources, as noted in the BlackCat catalog \cite{2016A&A...587A..61C}, where there are only four X-ray persistent BHs or candidates in the Milky Way, plus two sources where the nature of the compact object is ambiguous.  On the other hand, this catalog lists 59 transient BHs or candidates in the Milky Way, through 2014, while an additional 17 transients from MAXI and Swift have been suggested as BH candidates, using observations with the Neutron Star Interior Composition Explorer (NICER), from mid June 2017 through mid May 2024.

Targets of opportunity (ToOs) are also derived from major changes in the intensity and/or spectrum of known X-ray sources. WFM alerts for such opportunities will be issued in parallel with discoveries of new X-ray sources.  A few such examples are given here.  Accreting BHs and NSs have spectral states tied to structural changes in the accretion flow involving the accretion disk and episodes of activity in disk winds, a steady jet, or impulsive jets (Refs. \citenum{2006ARA&A..44...49R,2016LNP...905...65F,2023arXiv230405412N}).  Specialists in disk:jet, disk:corona, or disk:wind investigations need access to data collected at times when these components phase together, possibly reach temporary stability, and then evolve to a different state. Another example is the detection of coronal activity in nearby stars, a topic pertinent to both the physics of stellar coronae and also the environments of stellar activity in the context of exoplanet studies and habitable worlds. WFM detections of X-ray flares from nearby stars engages the capabilities of STROBE-X pointed instruments and also the diverse instruments of ground-based astronomy. WFM will also define ToOs from ground-based data analyses, where weekly sky maps can reach fainter classes of X-ray sources. Sky surveys from the burgeoning optical transients industry have demonstrated that some active galactic nuclei display strong variations that may be tied to accretion surges or tidal disruption events, some of which form strong jets that may be detected by the WFM \cite{2011Sci...333..203B}.

An overarching goal behind the WFM design is to provide data of maximum quality, so that the X-ray exposures during discovery and known-source monitoring will contribute vitally to astrophysical investigations. The defined data products and calibration goals reflect this perspective. The WFM archive plan is based on event lists, with full event information, to support user needs and creativity. On the other hand, X-ray light curves in several energy bands, and X-ray spectra for selected bright sources, will be routinely provided as a basis for many types of user programs. Illustrative scientific applications are as follows. For bright GRBs that are not chosen for immediate STROBE-X slew, WFM will continue to obtain 2-50 keV coverage as the prompt emission fades. This supports long-GRB modeling efforts by reducing the time gap between the prompt emission and the afterglow, while providing broad energy coverage to characterize the spectral evolution. For X-ray sources with intensity at 0.25 Crab (or higher), the WFM will provide quality X-ray spectra at a timescale of 1000 s (or faster) by performing the coded mask deconvolution in discrete energy bins, e.g. bins that oversample the SDD spectral resolution by a factor of three. Such efforts would provide 10--100 X-ray spectra each day, e.g., in 1000 s or 100 s exposures, depending on brightness. X-ray sources selected for this effort would include $\sim 15$ persistently bright X-ray binaries (mostly NS) and BH transients during bright intervals that often last from weeks to many months.

The detailed scientific goals of the STROBE-X Probe are described in Ref.~\cite{JATISOverview}. The WFM plays an essential role in each of the four scientific objectives of the Mission.  The goal to pursue multimessenger astrophysics via the electromagnetic counterparts of GWs and neutrino sources, depends, in part, on the WFM's capability to detect prompt emission (i.e., short GRBs) in time/location coincidence with a detection by GW facilities.  This effort expands on the ground-breaking cascade of results beginning with GW170817 and culminating in the evidence that the kilonova formed by the NS merger is a source of the heaviest metals that we see in the Universe \cite{2019Natur.574..497W}. The second objective, to characterize the nature of explosive transients, also relies on the WFM for recognition of the such events, which include ultra-long GRBs, subluminous GRBs, supernova breakouts, X-ray flashes, and other types of sudden X-ray flares. WFM has key roles in the overall effort to sort out the phenomenology and to identify progenitors and emission mechanisms. This is accomplished via source discovery, measurements of light curves and spectral evolution, and position localization that enables followup observations with contemporaneous facilities. For the third STROBE-X goal, to probe strong gravity and extreme physics, an important sub-theme is the determinations of spin values for black hole binaries (BHB), using multiple techniques, in the same outburst, for each target. The known BHB population is dominated by X-ray transients \cite{2016A&A...587A..61C}, and the WFM provides measurements of the intensity and spectral state for BHBs and candidates that can be observed with HEMA and LEMA at any time. The fourth goal is to capture the X-ray view of the "Dynamic Universe", with wide-field monitoring and broadband spectroscopy. This is more than the sum of science themes for different classes of transients, since it recognizes science process and discovery space, and it leads to requirements such as the timescales for WFM ToO alerts to the community (30 s to issue; 5 min to receipt), the slew rate of STROBE-X, and the timescale for uploading observing commands for a ToO selected from an external resource (24 hr). 

Estimates for the WFM annual yield of transients: 5 GW counterparts per year, 100 long GRBs; 100 extreme stellar flares; 400 Type I X-ray bursts (including ~15/year from sources not detected before or after the burst, several superbursts, 25 X-ray binaries (some recurrent), 5 jetted TDEs, 1 SGR/magnetar, and 15 transients in new or poorly defined classes (Refs. \citenum{2015ApJ...815..102F,2019arXiv190303035R,2004NuPhS.132..486I,2003A&A...411L.487I}).  Further considerations of WFM performance and comparisons with other missions are given in Section \ref{sect:performance}.

\section{Instrument Components}

%Figure 1

\begin{figure}[b]
    \centering
    \includegraphics[width=6.0in]{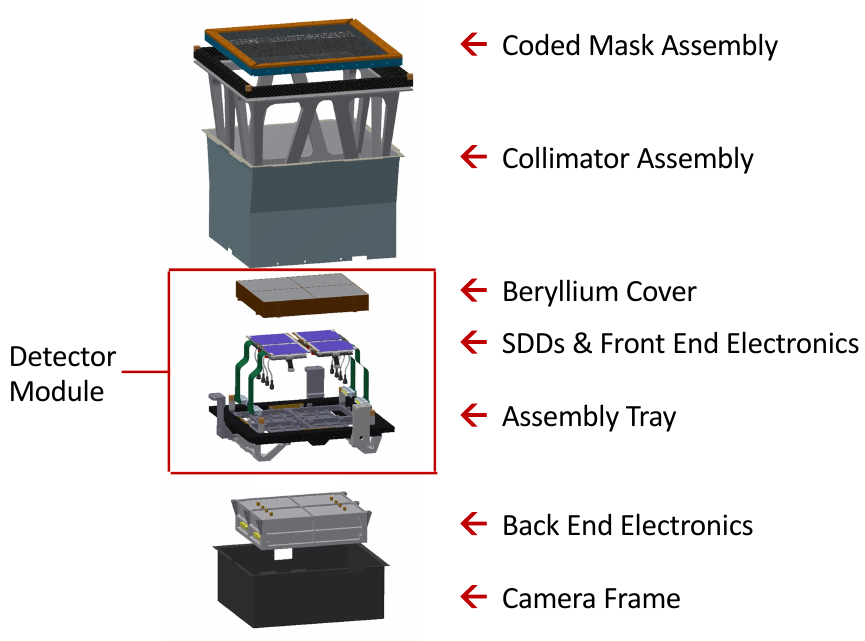}
    \caption{Illustration of the primary camera components of the WFM for STROBE-X}
    \label{fig:camera}
\end{figure}

The WFM camera components are shown in Fig.~\ref{fig:camera}, and the component functions are described below. In the plan for WFM-STROBE-X, hardware responsibilities include institutions in the US and Europe, and much of the WFM design heritage and expertise is engaged with the participation of WFM co-investigators and collaborators.  In brief summary, MIT (the principal investigator team) provides the coded masks and collimators, camera assembly, FPGA programming for the back end electronics, and detector/camera calibrations. The SRON Netherlands Institute for Space Research is the contract lead for the detector modules (see below), which are fabricated in partnership with the Institute of Space Sciences of the Spanish National Research Council (ICE-CSIC), and the National Institute for Nuclear Physics (INFN) in Italy. The US Naval Research Lab is producing the ASICs for the front end electronics boards, to read out the SDDs.  The Southwest Research Institute provides the hardware components for the back end electronics (including FPGAs and power supplies) and also the WFM Instrument Control Unit, which operates the instrument and conducts ToO searches. 

\subsection{Detectors}
The SDDs for both WFM and HEMA are produced in Italy by Fondazione Bruno Kessler (FBK) under contract with the National Institute for Nuclear Physics, Italy (INFN). The SDD lots for WFM and HEMA differ in anode pitch (169 $\mu$m vs. 970 $\mu$m), to increase spatial resolution for WFM imaging. A second difference is that the HEMA SDDs use a confined drift configuration, so that events stay on one anode.  Each WFM camera contains four SDD tiles, and 32 tiles is the in-flight total for the 8 WFM cameras.  The tile is 65 mm $\times$ 70 mm (active area), and the Si is 0.45 mm thick, providing energy response at 2--50 keV.  

Anode readouts lie along two sides of each tile, while high voltage ($-1300$ V) is applied along the tile’s center line, in between the arrays of anodes.  When an X-ray photon produces an electron cloud in the SDD, the electric field drives the electrons to the anodes. Positions for each X-ray event are determined as follows. The anode pitch is chosen to spatially resolve the charge cloud that is deposited after X-ray ionizations in the device. Each event is sampled with the anodes that best capture the integrated charge. The spatial profile of the collected charge, across these anodes, is centered to an accuracy of 60 $\mu$m \cite{2022SPIE12181E..1YH}. In the orthogonal direction, i.e., along the anodes, the event position is coarsely inferred from the FWHM of the event profile across the anodes. Events near the readouts have a narrow spatial profile, while events that originate near the center of the SDD have broad profiles, due to charge diffusion along the travel path from the event position to the anodes.  In the coarse-position direction the spatial resolution is 8 mm (FWHM). Each camera delivers X-ray counts that can be represented as a 2-dimensional array that corresponds to the 1082 x 9 position resolution elements of the detectors.  The coarse-resolution dimension separates the shadows of X-ray sources that happen to be aligned along the fine-resolution axis of the same camera.  Given the deployment of camera pairs with the same pointing direction, but with orthogonal fine and coarse axes, the coarse positioning capability does not significantly impact the accuracy of WFM localizations.  This is illustrated in Fig.~7 of Ref.~\citenum{2022SPIE12181E..1YH}.  However, coarse positions provide an increase by a factor of 9 in the number of independent detector measurements, and this reduces source confusion and enhances measurements of faint X-ray sources in crowded fields. Given the disparity in position resolution, per axis, single WFM cameras are sometimes described as "1.5-dimensional" imagers. In deployment, two co-pointing WFM cameras operate in tandem, with one rotated by 90$^\circ$ from the other, to measure a portion of the sky with symmetric 2D imaging.

X-ray events in the SDD are read out by the Front End Electronics (FEE), which were designed at SRON\cite{2022SPIE12181E..67Z}.  There is a close coupling between the four SDDs and the four FEEs contained in each camera. The SDD anodes are wire bonded to the readout ASICs in the FEE. This SDD/FEE ``sandwich'' enables the output stream of SDD analog signals per event, while also contributing mechanical strength. The FEEs for STROBE-X will use ASICs developed by the US Naval Research Lab\cite{JATISASIC}, with support from the APRA program. 

Each Detector Assembly (DA) includes the SDD/FEE sandwich with its mechanical structure, Cooling Plate (CP) and Invar Bracket (IB), a thermal strap and an output harness. The four DAs are mounted in a detector tray, which is further joined to a cold plate and placed below a protective  cover (beryllium or plastic) for orbital debris and micrometeorites. The sum of these components constitutes a detector module (see Fig.~\ref{fig:camera}).

In the current plan, the detector modules (one per camera) will be provided by SRON. After receipt of the SDDs from Italy, SRON will wire bond each SDD tile to its FEE PCB, and these ``sandwiches'' are attached to mechanical parts (CP and IB) to form a detector assembly (DA). The DAs are sent to the WFM team at ICE-CSIC in Spain, where they will be mounted and aligned, four per camera. The four DAs and a protective attached cover are then attached to the Detector Support Plate, completing the build of the so-called Detector Module.  Detector modules are sent back to SRON for final tests, and then they will be delivered to MIT. The SRON, ICE-CSIC, and INFN principals have been working together for more than a decade on WFM designs and implementation.  

\subsection{Detector Electronics}
Further details are given for the FEE, which was introduced in the previous subsection. Functions of the FEEs and its interfaces with the Back End Electronics (BEE) are described in Ref.~\citenum{2022SPIE12181E..67Z}.
Basically the FEEs supply power to the SDDs and ASICs, they operate survival heaters for the SDDs, and they deliver analog signals for events and housekeeping data to the BEE. The primary functions of WFM electronics packages are summarized in Table~\ref{tab:efunctions}. 

\begin{table}[htbp]
    \centering
        \caption{Primary functions of the WFM electronics packages}
    \label{tab:efunctions}
    {\footnotesize
    \begin{tabular}{llll}
    \hline
         &  Front End Electronics (FEE) & Back End Electronics (BEE) & Instrument Control Unit (ICU) \\
         \hline
         Lead Inst. & SRON & MIT/SwRI (initial design DLR) & SwRI (Initial design DTU/Sabanci U)\\
         Quantity & 1 per det. (4/camera) & 1 per camera & 2 (cold spare; 1 per 8 cameras) \\
         \hline
         \multicolumn{4}{c}{Functions} \\
         \hline
         & Bias voltages to SDD & Power (LV/MV/HV) to FEE & Power distribution \\
         & Power \& configure SDDs & ADC conversion & Camera state commanding \\
         & Power \& configure ASICs & Event time tags & Data handling unit \\
         & Read SDD signals (ASICs) & Filter events & Memory (data; coded mask libraries)\\
         & Send events to BEE & Determine event positions & Send PPS time synch to BEE \\
         & Send kousekeeping to BEE & Write housekeeping packets & Online transient detect/alert \\
         & Survival heaters & Write event packets & Online transient slew request \\
         & Mechanical support to SDD & Send packets to ICU & Interface to Spacecraft \\
         \hline
    \end{tabular}
     }
\end{table}

Originally, the FEEs were designed with IDeF-X ASICs (32-channels each) in France\cite{2022ITNS...69..620B}. In 2023, the US Naval Research Lab began to develop the readout ASICs for the FEEs in both the WFM and HEMA, with funding from the NASA-APRA program.  The NRL ASICs for WFM have 64-channels, with six devices per side, to sample 384 anodes (i.e., 65 mm SDD length with 169 $\mu$m pitch). SRON is planning the accommodation of the NRL ASICs for the FEEs for STROBE-X.

In each camera, the signals captured from triggered events are sent from the FEEs to the BEE for further processing. The anode signals are digitized and time tagged, and the multi-anode data is converted to event positions and total amplitude. Filtering is then done to separate the good events from other categories, such as high amplitude events caused by energetic particles. Additional functions of the BEE include preparation of telemetry packets for events and housekeeping data, supplying power at high and low voltage to FEEs, and sending packets and data to the Instrument Control Unit (ICU) for telemetry and also for onboard ToO searches. 

The original design of the BEE for LOFT and revisions for eXTP were done by the WFM group in T\"{u}bingen, Germany, supported by DLR.  For WFM-STROBE-X, the BEE functions are the responsibility of MIT, and SwRI will control the hardware selections and PCB manufacturing. These choices are motivated by the idea that the PI team (MIT) should have direct control of in-flight event selections and on-board calibrations for position and photon energy, to optimize both calibration maintenance and the speed of responses to unanticipated issues that may be encountered during the operation of a long-term mission. Members of the T\"{u}bingen group remain active hardware participants for HEMA\cite{JATISHEMA} and expert consultants for WFM-STROBE-X. For the BEE boards, SwRI will select the commercial space-qualified ADCs, the power supplies (HV, LV, and MV), and FPGA hardware.

\subsection{Coded Masks}
The angular viewing properties of WFM cameras are determined by sizes of the detector surface and the coded mask, and the height of the mask above the detector plane (see Fig.~\ref{fig:geometry}).  For a mask width, $m = 260$ mm, mounted at a height $h = 202.9$ mm above the SDDs, which have active area width, $w = 143.5$ mm (including a central gap between SDD tiles), the half-angle for viewing an X-ray source with a fully coded mask shadow on the detector plane is $\tan^{-1} ((m/2 - w/2) / h) = \pm 16.0^\circ$.  The offset angle reaching one-half mask coverage is $\tan^{-1} (0.5 m / h)) = \pm 32.6^\circ$,  and the limit for zero coding is $\tan^{-1} (m + w) / 2h) = \pm 45.2^\circ$.

%Figure 2
%% Camera drawing is here

\begin{figure}[htbp]
    \centering
    \includegraphics[width=3.0in]{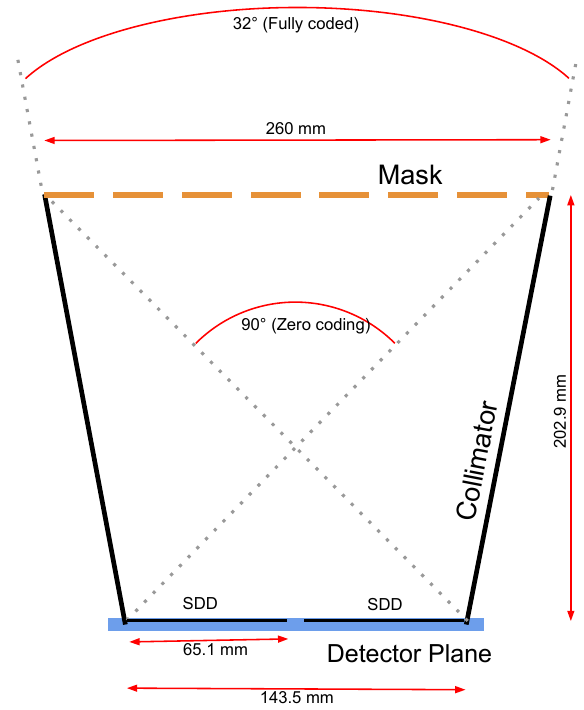}
    \caption{WFM mask and detector geometry, showing the fully-coded and zero-coded field of view angles.}
    \label{fig:geometry}
\end{figure}

The mask material, which must be opaque to X-rays out to 50 keV, is Tungsten, with a thickness of 150 $\mu$ m.  This a persistent feature in the WFM designs for LOFT and eXTP \cite{2012SPIE.8443E..5PE,2022SPIE12181E..1YH}.} However, the mask pattern and open fraction will be adjusted for STROBE-X, compared to previous designs. For STROBE-X, the nominal plan for the coded mask reserves 10\% of the area for strips to stiffen the mask, while the remainder is divided between randomly open and closed elements. This yields a mask that is 45\% open, and this higher value, compared to the LOFT design, takes advantage of the high telemetry capacity of STROBE-X to gain higher count rates for transients that are fast and bright.  The sizes of the mask elements are nominally 250 $\mu$m $\times$ 32 mm, or a factor of 4 larger than the spatial resolution along each axis in the detector plane. In between the solid stiffening strips, each row in the mask pattern must be minimally correlated with the other rows, to prevent aliases that can skew the identification of transients. The final selections for the mask patterns will be completed during STROBE-X Phase A.

Nominally, many mask edges (i.e., open/closed boundaries) will be measured along the camera’s fine-axis, where the resolution of 60 $\mu$m projects to 1 arcmin on the sky, for a mask height of 202.9 mm. This assures consistency with the WFM goal to localize new transients to a position uncertainty of 1 arcmin. Qualitatively, where statistics are inadequate to accurately constrain the mask edges in the detector plane, the localization accuracy reverts to the size of single mask elements, and 250 $\mu$m corresponds to 4 arcmin. Simulations of the WFM instrument indicate that a 1' x 1' localization square is achieved for a camera pair when the X-ray source is detected with a signal-to-noise of 10 \cite{2012SPIE.8443E..5PE}.

\subsection{Instrument Control Unit}
The ICU, to be provided by SwRI, has high flight heritage derived from the Central Instrument Data Processor (CIDP) payload suite computer on the Magnetospheric Multiscale mission (MMS). The electronics box is a 6U extended cPCI Chassis enclosure, behind 125 mils of Aluminum. Two identical units will be delivered, and one is a cold spare. 

The ICU design adopts the functional requirements of the ICU for WFM-\textit{eXTP}\cite{2022SPIE12181E..66K}. There is a Power Analog Board, a Digital Board, and a Memory/Trigger Board.  The ICU interfaces with 8 WFM cameras, via their BEEs, to provide power distribution, camera commanding, data handling, memory storage (telemetry, science, and mask response libraries), and time synchronization pulses.  The ICU processes realtime data for each camera pair to perform several types of online transient searches, as has been pioneered by the \textit{Swift} mission \cite{2013ApJS..209...14K}, in operation for SVOM/ECLAIRs \cite{2019MmSAI..90..267S}, and planned for WFM-\textit{LOFT} and WFM-\textit{eXTP}\cite{2022SPIE12181E..66K}. 

The ICU interfaces with the Spacecraft (S/C) to relay commands to the cameras and to provide time synchronization pulses (PPS) to the BEEs, to be used to time-tag events.  The S/C interface also provides to the ICU a STROBE-X pointing schedule and orbit ephemerides to assist in the assessment of feasibility to submit slew requests to the S/C for high-priority ToOs. On the output side, the ICU sends telemetry packets to the S/C, along with ToO alerts to the science community and for S/C slew consideration.

The S/C electrical interface utilizes a redundant RS422 UART for commanding, telemetry, and PPS, and a LVDS High Speed Serial interface for transmission of science data directly to the redundant S/C Solid State Recorders. Flight Software (FSW) executes on a LEON3-FT dual core processor on the Digital Board,  running the Real-Time Executive for Multiprocess Systems (RTEMS) operating system.  FSW executes S/C commands to the cameras and transmits telemetry, and provides camera control, telemetry, and operational modes for all 8 cameras. The FSW uses a modular, layered architecture that benefits from 85\% reusable components from the successful space missions, MMS and Cyclone Global Navigation Satellite System (CYGNSS).  The FSW will adapt interface software already developed for RAL and PUNCH cameras. The architecture is based on a mature software bus that distributes messages with low latency.  Critical tasks are run synchronously with real-time checks to ensure deadlines are satisfied.  

\section{WFM Operations on STROBE-X}

\subsection{Orbit and Space Environment}

STROBE-X will operate in a circular orbit, with altitude between 550 and 575 km and an inclination angle $< 15^\circ$ \cite{JATISOverview}. This orbit provides relative protection against the radiation environment, which can degrade SDD energy resolution over time. During the 5-year prime mission, the maintenance of WFM energy resolution may be assisted by operating the SDDs at the lower end of the temperature range (which is -20 to -40C) and by heat-annealing the SDDs, as will be done for the HEMA instrument \cite{JATISHEMA}.

\subsection{Camera Mounting on the Spacecraft}

\begin{figure}[b]
    \centering
    \includegraphics[width=5.0in]{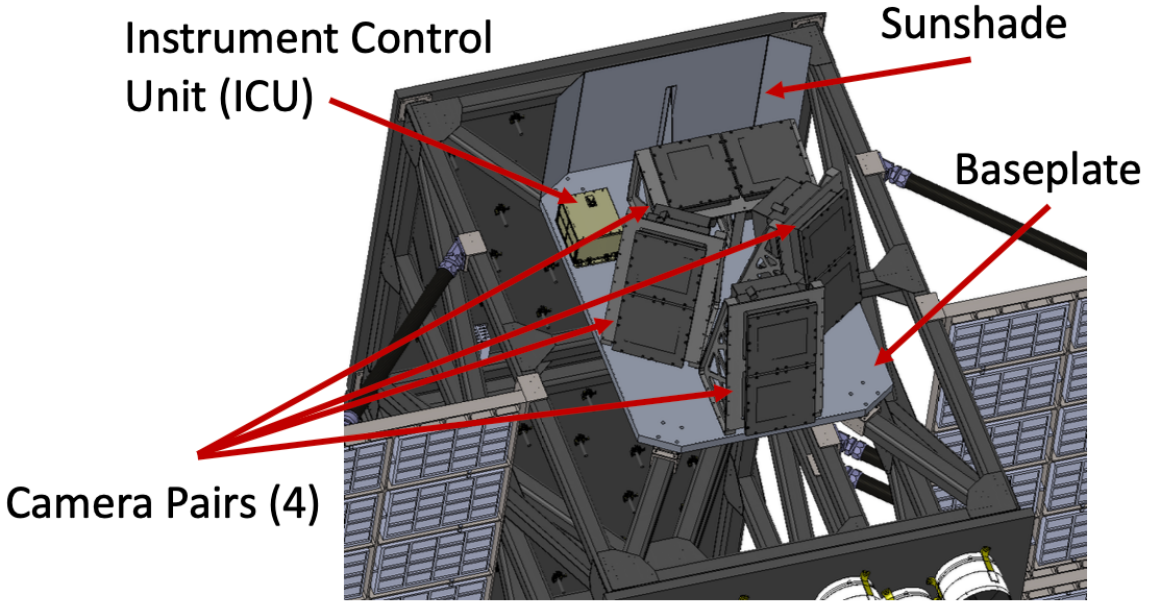}
    \caption{Rendering of the STROBE-X observatory, showing the WFM mounted on a plate at the top of the Spacecraft.}
    \label{fig:mount}
\end{figure}

The WFM cameras are designed with a minimum Sun angle of 45$^\circ$, matching the requirement for instruments HEMA and LEMA.  In addition, WFM cameras must be shaded from the Sun during normal operations, to maintain proper operating temperatures in the SDDs (i.e., -20 to -40 C). There is a further requirement to avoid sunlight on any portions of the coded masks during observations, since mask expansion can skew the positions of the mask shadows, compared to models. For STROBE-X, the mounting solution is simplified by the convention that the spacecraft will select the roll angle that keeps the Sun in the ($x, -z$) plane, where $x$ is the pointing axis for HEMA and LEMA, and $-z$ is the long axis of the spacecraft in the direction of the launch vehicle. The four WFM camera pairs are mounted on a plate at the top of the optical bench (Fig.~\ref{fig:mount}), i.e., in the $+z$ above the optical bench, where the base plate and spacecraft body keep the WFM in Sun shade for most of the allowed range of the target's Sun angle. 

% Figure 3
% camera mounting on STROBE-X

However, adjustments must be made for two remaining Sun issues. First, the spacecraft body does not shadow the WFM cameras when the angle ($\phi$) between the target and the Sun is in the range $150 < \phi < 180^\circ$.  A Sun shade is attached to the WFM mounting plate on the $-x$ side of the WFM (Fig.~\ref{fig:mount}) completing the solar shade requirement for the full range of allowed Sun angles: $45 < \phi < 180^\circ$. Second, camera pair \#1 would nominally point in the direction of the STROBE-X target ($+x$ axis), but the Sun would then shine on the cameras’ masks when the Sun angle is near the minimum value ($\phi = 45^\circ$).  To keep the Sun off the masks, the cameras are pulled back from the edge of the mounting plate, and the cameras are tilted up by $15^\circ$ toward $+z$.  In this position, the STROBE-X target is in the fully coded mask area of camera pair \#1, while the FOVs are free from occultation by the edge of the base plate. Camera pairs \#2 and \#3 are kept in the same tilted plane, but rotated by $65^\circ$, on either side of the central pair (\#1). This offset is twice the cameras’ half-coded half angles, i.e., half-coded mask lines abut each other, for camera pairs \#1 with \#2, and \#1 with \#3, so that the net open fraction is roughly flat between the camera pair pointings.  Camera pair \#4 is offset by $65^\circ$ in the $+z$ direction, relative to \#1.  The camera pair pointing directions are then (polar coordinates, $\theta, \phi$, with $\theta = 0$ along $+z$, and $\phi = 0$ along $+x$): (75.0$^\circ$, 0.0$^\circ$), (83.72$^\circ$, 65.75$^\circ$), (83.72$^\circ$, 294.25$^\circ$), and (10.0$^\circ$, 0.00$^\circ$). 

% Figure 4
% WFM sky view

\begin{figure}[b]
    \centering
    \includegraphics[width=5.0in]{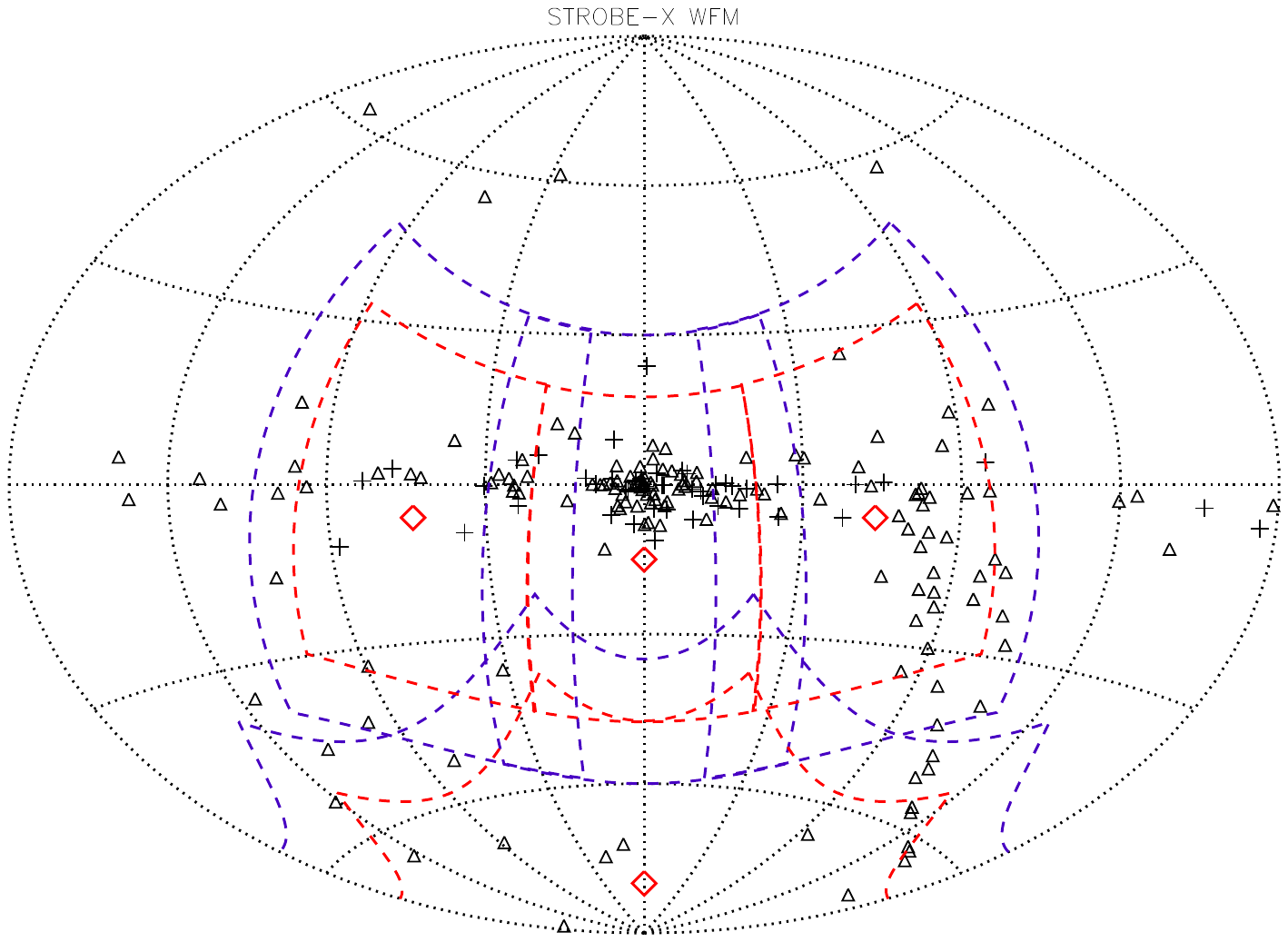}
    \caption{WFM field of view, shown in galactic coordinates and mapped with an Aitoff projection. Red diamonds mark the pointing centers for each camera pair, which have half-coded (zero-coded) boundaries traced with dashed red (blue) lines.
    A small black "+" shows locations of 46 persistent X-ray sources, while the small triangles show locations of 151 X-ray transients with bright maxima. Other types of sources, such as long Gamma Ray Bursts, electromagnetic counterparts of neutron star mergers, and nearby AGN that may undergo outbursts, are distributed isotropically.}
    \label{fig:fov}
\end{figure}

The FoV pattern on the sky for the WFM-STROBE-X mounting configuration is illustrated in Fig.~\ref{fig:fov}.  Here is it convenient to use galactic coordinates (Aitoff projection), to point STROBE-X at $(l,b)$ = (0.0, 0.0), and to choose the S/C $z$ axis to align with the North Galactic Pole.  The four camera pair pointing are then ($l, b$): (0.0$^\circ$, $-15.0^\circ$), ($65.75^\circ$, $-6.28^\circ$), ($294.25^\circ$, $-6.28^\circ$), and (0.0$^\circ$, $-80^\circ$). The alignment of adjacent camera pairs at the half-coded lines (red dashes) is evident, while the FoVs to the zero coded limit is also shown (blue dashes). The locations of 46 persistent X-ray sources with intensity (2-20 keV) above 10 mCrab is shown with the ``+'' symbol, while 151 X-ray transients with bright maxima (some recurrent on timescales from 2-50 years) are plotted with open triangles.

\subsection{Sky Coverage}
The four WFM camera pairs have fixed mounting directions, and overall sky coverage is gained when STROBE-X points at different targets.  Operating in low Earth orbit, STROBE-X is expected to slew many times per day, in part to avoid 
losing time to Earth occultations that would generally interrupt exposure times longer than a spacecraft orbit ($\sim 90$ min). Since the instrument FoV is already one-third of the sky in a single pointing, an accumulation of 30 well-separated targets per day (i.e., two per orbit) would produce roughly five exposure intervals for any point in the WFM field of regard, after consideration of Earth occultations in the WFM FoV. 

The rule that no WFM camera pair can be pointed to within a $45^\circ$ cone centered on the Sun does not necessarily imply
that no WFM exposure can occur within that same cone.  In the limiting case where the target has a sun angle of $45^\circ$, 
camera pair \#1 has a pointing direction that is closest to the Sun, given the practice to choose a spacecraft roll angle to keep the Sun in the $x,z$ plane. Given the $15^\circ$ offset in the camera \#1 pointing direction, relative to the target direction, the $45^\circ$ FoV edge, to zero coding, extends to a point that is $30^\circ$ from the Sun.  Thus, a series of targets with $45^\circ$ Sun Angle and different spacecraft roll angles would defines the WFM field of regard, which only excludes exposures within $30^\circ$ of the Sun. The mask-shadow-weighted exposures near this limit would be quite small, but loss of contact with a bright transient in the ecliptic plane is formally limited to 2 months of the year, rather than three.

\subsection{Data Collection and Archive}
\label{sect:data}  % \label{} allows reference to this section

The WFM normally conducts observations in ``event mode'', in which the primary information for all good events is telemetered to the science archive.  The bit allocation for WFM event mode is as follows. Each of the eight WFM cameras is implicitly  distinguished by the Application ID for its event packets.  The event words themselves are given the following budget: SDD tile ID (2 bits), anode side (1 bit), digitized pulse amplitude (10 bits; 0-60 keV in 60 eV steps), fine position value (12 bits; 65 mm / 60 $\mu$m resolution oversampled $\times$ 3.8), coarse position value (5 bits; 70 mm / 8 mm oversampled $\times$ 3.7), event time (12 bits ; see below), and event flags (3 bits). The total event word size is then 45 bits.

As with the other STROBE-X instruments (and also NICER), accurate timing knowledge is maintained by transmitting a PPS signal and the corresponding GPS time from the spacecraft to the BEE, which time tags each event.  The tick count from the BEE clock (18 MHz) is captured at each PPS signal, while the tick count corresponding to the last GPS signal is recorded in the event packet headers. CCSDS packets for each WFM camera are issued for 100 events or 0.1 s, whichever comes first. The assignment of 12 bits for event time thus conveys a time resolution of 0.1 s / 4096 = 24 $\mu$s. It has been demonstrated with NICER that this timing system can be calibrated to an absolute accuracy of 300 ns, and so the calibrated WFM time tags in the archive will have absolute accuracy as good as the time resolution.

The WFM background rate (2-50 keV) is calculated to be 480 c/s/camera (assuming the mask open fraction is 0.45), and the count for the Crab Nebula is 188 c/s/camera. Based on the average all-sky intensity of discrete X-ray sources (21 Crab) in the survey by the All-Sky Monitor of the Rossi X-ray Timing Explorer (1996-2012), we estimate that the WFM will produce an average of ~7000 X-ray events per second. This number includes corrections for field-of-view effects and Earth occultations.  Considering an assignment of 44 bits per event and a 30\% overhead for event mode telemetry, the average telemetry rate is 400 kbps. Margin is gained by not considering operating losses due to the South Atlantic Anomaly or to STROBE-X slews. Allowances for WFM housekeeping packets are not comparatively significant. At maximum, the WFM telemetry rate could increase by a factor of 2, for a few hours per day, in the case of a rare 10 Crab transient and the condition that both Sco X-1 (11 Crab) and this transient are in the fully coded region of any cameras.

\section{Instrument Calibration}
All of the SDDs are initially calibrated on the ground with the Modulated X-ray Source (MXS)\cite{2016SPIE.9905E..4WL,2016SPIE.9905E..1HG}. This device features a rapidly pulsed (600 Hz) electron beam that impinges on a metal target, producing a pulsed X-ray fluorescence source at the photon energies tied to the target material (Al 1.5 keV; Ti 4.5 keV ; Cu 8.0 keV, etc.) Tests with the MXS will measure, over discrete values in photon energy, each detector’s spatial resolution, flatness, energy resolution, and relative QE. In addition, absolute detector calibrations will be sought for a couple of detector modules at the Bessy-II electron storage ring in Berlin, as was performed for NICER SDDs \cite{2017HEAD...1610410P}.  Such tests measure the event trigger efficiency and absolute QE versus photon energy.

Full camera calibrations are conducted in space.  Measurements of the camera boresights, the coded mask alignments, and the final throughput values through the solar blanket are derived from observations, with each camera pair, of  Sco X-1, the Crab Nebula, and other X-ray sources. These tests will be performed during in-orbit checkout, or soon afterwards if calibration targets are unavailable, due to Sun constraints.

\section{Instrument Performance}
\label{sect:performance}  % \label{} allows reference to this section

\subsection{Performance Summary}
Instrument performance specifications for the WFM are summarized in Table \ref{tab:wfmspecs}. Many of these parameters are tied to details given in previous sections. while count rates and sensitivity limits were derived from simulations conducted at SRON.

\begin{table}[htbp]
    \centering
        \caption{WFM Instrument Performance}
    \label{tab:wfmspecs}
    {\footnotesize
    \begin{tabular}{lll}
    \hline
    Energy Range & 2–50 keV & 10\% limits in SDD at 1.45 keV, 33.4 keV \\
    Effective Area & 364 cm$^2$ & per camera pair, no mask \\
    Energy Resolution & 300 eV & at 6 keV \\
    Time Resolution & 24 $\mu$s & limit is time bits per event \\ \\
    Field of View & $\pm 16^\circ$ ; $\pm 32^\circ$ ; $\pm 45^\circ$  & fully coded ; half coded ; zero coded limit \\
    Sky coverage, 4 camera pairs & 4 sr ; 7 sr & half coded ; zero coded limit \\
    Angular Resolution & $4' x 2.3\circ ; 4' x 4'$ & projected mask element for camers ; camera pair \\  
    Source Localization & 1 arcmin & source detections $> 10 \sigma$ \\
    Source Localization & 4 arcmin & sources near detection limit \\ \\
    Crab Nebula count rate & 376 c/s & full band ; camera pair ; 0.45 open mask \\
    Background count rate & 960 c/s & full band ; camera pair ; 0.45 open mask \\
    Sensitivity in 1 s & 920 mCrab & $5 \sigma$ \\ 
    Sensitivity over days & 4 mCrab & 5$\sigma$ ; 50 ks\\ 
    \hline
    \end{tabular}
     }
\end{table}

\subsection{Deep Sky Maps and Source Confusion}

The WFM has the capability to handle a large number of X-ray sources.  The instrument provides a capacity to make deep sky maps, even in the relatively crowded region near the Galactic Center. First, the WFM confusion cell on the sky is limited to pairs of sources with angular separation of only 1-5 arcmin, depending on source intensities, given the crossed lines of fine positions provided by the two cameras in a given pair. Second, each camera pair provides a large number of position resolution elements, which, in principle, corresponds to the number of independent measurements available to handle the sources and background in the FoV. The active area, per SDD tile, is $64.9 \times 70.0$ mm, while the position resolution is $60 \mu m \times 8$ mm.  Each camera contains 4 tiles, yielding  4 x 1082 x 9 spatial resolution elements. The full FoV for each pair of cameras thus provides more than 77,000 independent measurements. 

The expected number of X-ray source, down to a given intensity level, can be can be estimated from recent sky surveys.
The new SRG/eROSITA all-sky survey\cite{2024A&A...682A..34M}, examined in the 2.3-5 keV energy band, predict 226 sources above 2 mCrab in $90^\circ \times 90^\circ$ square centered on the Galactic Center, extrapolating from the 113 sources cataloged in the Western half that is within its purview. The task to find 226 unknown intensities using 77,000 independent measurements is certainly feasible and implies a capacity to go far deeper with very long exposures.  The sensitivity level of 2 mCrab roughly corresponds to a week of WFM exposures.  The tradeoff with this narrow-field, sensitive, scanning instrument is that it takes SRG/eROSITA six months to complete a sky scan.

Finally, the practice of catalog-driven analyses for coded mask data is worth mention.  The convention is to use precise (e.g., 1 arcsec) positions of X-ray sources to model the mask shadows, and that is why the only variables in the imaging analyses, in a given energy band, are the source intensities. New transients are appended to the analysis catalog, as soon as their positions are accurately determined, and their light cures are considered final after the deconvolution modeling is iterated. The community demand to obtain a very large number of source light curves, the majority of which may pertain to transients in a dormant state, is satisfied without concerns for source confusion or model overload by first defining a minimum analysis catalog, e.g., for sources detected at 90\% confidence in moderately long exposures, and then adding the "dead" sources, in small and well-separated batches, to the minimum catalog to produce the desired data products. 

\subsection{Comparison with Other Instruments}
Wide-angle instruments are sometimes compared with a figure of merit known as, "grasp", which is the product of the full FoV (sr) and the effective area (cm$^2$), displayed as a function of photon energy.  The effective area curve is the response curve calculated with considerations for the detector and any blankets or covers in the design.  For coded mask instruments there are two additional factors, the open fraction in the coded mask, and the average mask fraction over the instrument FoV, scaled to the value when fully coded.  The latter quantity has a higher value when the coded mask area is larger than the detector area, as occurs with the WFM (see Fig.~\ref{fig:geometry}).  

Table~\ref{tab:wideX} contains the quantities and references used to compare the grasp of the WFM with previous wide-angle instruments that operate in the conventional X-ray band, or roughly 1-15 keV. Since the grasp parameter does not consider the instrument background rate, which is a limiting factor for the measurement of faint sources, the monitoring sensitivities are also given in Table~\ref{tab:wideX}.  Sensitivities are given at 5$\sigma$, averaged over the field of regard (excludes Sun avoidance), for one day of exposure time, in the corresponding instrument's energy band, in units of mCrab, where 1 mCrab is equivalent to 2.1$\times 10^{-11}$ erg cm${-2}$ s$^-1$ at 2-10 keV. Notable, operating monitors missing from Table~\ref{tab:wideX} with effective area centered on hard or soft X-rays, are the Swift-BAT instrument \cite{2004ApJ...611.1005G} (15-150 keV), SVOM/Eclairs \cite{2016arXiv161006892W} (4-150 keV), and the Wide Field X-ray Telescope on the Einstein Probe (https://ep.bao.ac.cn ; 0.5-4 keV).

\begin{table}[htbp]
    \centering
        \caption{Grasp and Sensitivity for Wide-Angle Instruments in the Conventional X-ray Band}
    \label{tab:wideX}
    {\footnotesize
    \begin{tabular}{lccccccl}
    \hline
    Instrument/ & FOV (sr)  & Range  & Sensitivity$^1$ & Mask Open & Transmission & Maximum Eff. & Response Curve \\
    Spacecraft  & (to zero) & (keV)  &     (mCrab)     & Fraction  & (FoV avg.)   & Area (cm$^2$) & Reference \\
    \hline
    ASM/RXTE       &  1.0   & 1.5-12 &  30          &  0.50     & 0.301        &  59.4 (4.25 keV) & Ref.~\citenum{1996ApJ...469L..33L}   \\
    WFC/Beppo-SAX  &  0.90  & 2-30   &  16          &  0.25     &  0.25        &  140 (10.4) keV  &  Ref.~\citenum{1997AASup..122..299B}  \\
    MAXI/ISS       &  0.10  & 2-30   &  28          &  ...      &  0.50        &  267 (8 keV)     &   https://ep.bao.ac.cn \\
    WFM/STROBE-X   &  7.0   & 2-50   & 5.5          &  0.45     & 0.406        &  364 (8.7 keV)   &  https://strobe-x.org   \\
    \hline
    $^1$ Monitoring sensitivity, 5$\sigma$, averaged over field of regard, per day
    \end{tabular}
     }
\end{table}

\begin{figure}[htb]
    \centering
    \includegraphics[width=4.0in]{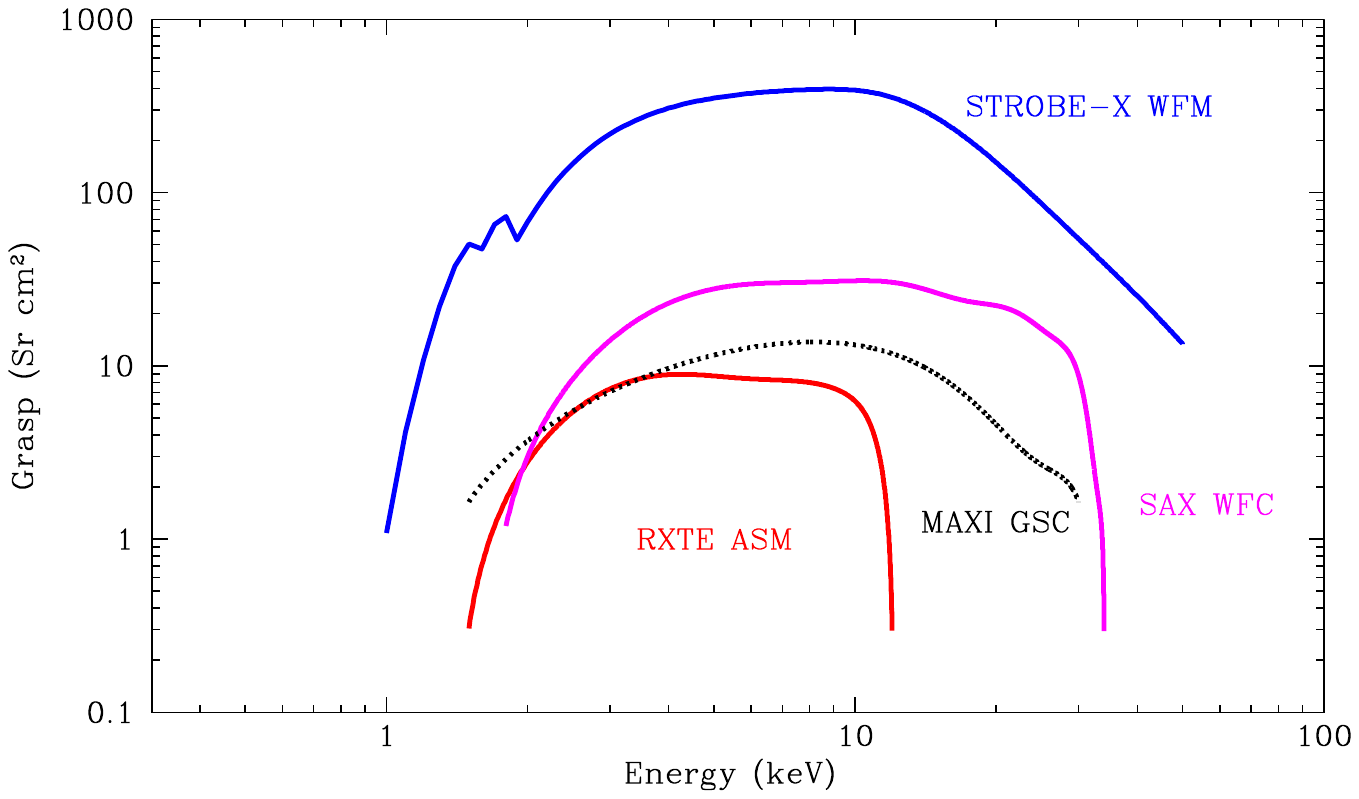}
    \caption{Grasp is the effective area curve times the solid angle of the instrument FoV, with the latter quantity weighted by the transmission factor at each point in the FoV. The WFM has substantially larger grasp compared to other wide-angle cameras with sensitivity centered on medium-band X-rays. A solid line is used for coded mask cameras, and a dotted line is used for the MAXI Slat Collimator.}
    \label{fig:grasp}
\end{figure}

Fig.\ref{fig:grasp} shows that the grasp of the STROBE-X WFM exceeds that of current and former wide-field instruments in the conventional energy band, with improvement by an order of magnitude.  The all-sky sensitivity is the best of this class. User participation and creativity are facilitated by the delivery of the WFM data in event mode, the policy to make all WFM data public, and the plan to deliver standard data products, including multi-band light curves and bright-source spectra. WFM transient discoveries and targets of opportunity determined from known sources will be the primary supplier of targets for the giant LEMA and HEMA instruments that point and deliver X-ray events with an unprecedented combination of high count rate, timing precision, and energy range.

\section{Summary}
  
The WFM is a system of coded mask cameras that will capture the continuously evolving X-ray sky with unprecedented continuity and data quality, providing a window to diverse classes of high-energy transients. For STROBE-X, four pairs of cameras instantaneously cover one third of sky to the limit of 50\% mask coding, and 56\% of the sky to zero coding. Camera boresights are fixed, and sky coverage is gained as STROBE-X is repositioned to observe many targets per day. The WFM energy range is 2-50 keV, energy resolution is better than 300 eV at 6 keV, every event is sent to the ground with a time resolution of 24 $\mu$ s. X-ray transients are localized to 1 arcmin when the detection significance is 10 $\sigma$. Onboard processing in the WFM Instrument Control Unit provides community alerts within five minutes of detection, for new transients and for prioritized types of variability in known sources.  Some alerts will trigger autonomous slews to engage the narrow-field instruments, HEMA and LEMA. The WFM plays essential roles for each of the primary objectives of STROBE-X.  For example, WFM detection and localization of short GRBs is a key to the multi-messenger goal to study the electromagnetic counterparts of gravitational wave sources involving a neutron star in a compact object merger. Another example concerns the capacity of HEMA and LEMA to investigate black hole binaries by bringing time domain astrophysics to the innermost stable circular orbit (ISCO), accumulation 200 to 2000 counts per typical dynamical ISCO circulations at 500 Hz (i.e., count rates of $10^5 - 10^6$ during many outburst maxima).  For this enterprise, the WFM defines the times and context for all the best targets. 

The central task of a coded mask camera is to record the position and photon energy of each event in the detector plane, so that the accumulated counts array can be deconvolved as a superposition of a set of offset mask shadows, each one linked to an X-ray source in the camera's field of view. The WFM detectors use silicon drift detectors (SDD) from FBK (Italy) are used, with four SDD tiles (each 65 x 70 mm) operating in each camera. The tiles have 384 anodes along two opposing sides, with high voltage applied across the center line. WFM cameras are described as "1.5 dimensional", since each camera has fine (4.3 arcmin on axis) and coarse ($2.3^\circ$ ) angular resolution, by virtue of the different detector resolutions along the anode and perpendicular axes. Pairs of cameras are deployed per pointing direction, with one camera rotated by $90^\circ$ to apply fine-positioning constraints to both axes.  Each SDD tile is mated to a Front End Electronics board that contains two rows of six 64-channel ASICs designed to read signals from any anode that is triggered for measurement. Power supplies, survival heaters, and other functions of the FEE follow conventional practices for space missions. The output signals and housekeeping information from the FEE boards are sent to the Back End Electronics (BEE, one per camera) for digital conversion, event screening, position determination in the detector plain, and telemetry formatting. Output data from the BEE are sent to the Instrument Control Unit, which has processes to relay the telemetry packets and to stage realtime data to search for different types of ToOs. The STROBE-X WFM instrument team brings forward a very strong design heritage, evolving from the 2013 Concept Study Report for LOFT for the European Space Agency.

\clearpage

\appendix    % this command starts appendixes

% \disclosures 
% \subsection*{Disclosures}
% The authors have no potential conflicts of interest to disclose.

% \subsection* {Code, Data, and Materials Availability} 
% Data sharing is not applicable to this article, as no new data were created or analyzed.

% \subsection*{Acknowledgments}
% Portions of this work performed at MIT were supported by NASA grant 80NSSC19K1287.

%%%%% References %%%%%

\bibliography{report_2}   % bibliography data in report_2.bib
\bibliographystyle{spiejour}   % makes bibtex use spiejour.bst

% Biographies Section omitted
%\listoffigures
%\listoftables

\end{spacing}

\end{document}